# Order out of chaos: Shifting paradigm of convective turbulence

Sergej Zilitinkevich,[a,b,c] Evgeny Kadantsev,[a,b] Irina Repina,[d,e,f] Evgeny Mortikov,[e,f,g] Andrey Glazunov[f,g]

[a] *Institute for Atmospheric and Earth System Research / Physics, Faculty of Science, University of Helsinki, 00014, Finland*

[b] *Finnish Meteorological Institute, Helsinki, 00101, Finland*

[c] *Faculty of Geography Moscow University, Russia*

[d] *Obukhov Institute of Atmospheric Physics, Russian Academy of Sciences, Moscow, 119017, Russia*

[e] *Moscow Center for Fundamental and Applied Mathematics, Moscow, 119991, Russia*

[f] *Research Computing Center of Lomonosov Moscow State University, Moscow, 119991, Russia*

[g] *Marchuk Institute of Numerical Mathematics, Russian Academy of Sciences, Moscow, 119333, Russia*

*Corresponding author*: Evgeny Kadantsev, evgeny.kadantsev@helsinki.fi

**Abstract**

Turbulence is ever produced in the low-viscosity/large-scale fluid flows by the velocity shears and, in unstable stratification, by buoyancy forces. It is commonly believed that both mechanisms produce the same type of chaotic motions, namely, the eddies breaking down into smaller ones and producing direct cascade of turbulent kinetic energy and other properties from large to small scales towards viscous dissipation. The conventional theory based on this vision yields a plausible picture of vertical mixing and remains in use since the middle of the 20$^{th}$ century in spite of increasing evidence of the fallacy of almost all other predictions. This paper reveals that in fact buoyancy produces chaotic vertical plumes, merging into larger ones and producing an inverse cascade towards their conversion into the self-organized regular motions. Herein, the velocity shears produce usual eddies spreading in all directions and making the direct cascade. This new paradigm is demonstrated and proved empirically; so, the paper launches a comprehensive revision of the theory of unstably stratified turbulence and its numerous geophysical or astrophysical applications.

# 1. Introduction

The concept of turbulence ("turba aliena") as chaotic motions of fluid elements ("atoms") was known already to Roman philosopher Lucretius ('De Rerum Natura'). He defined it as "aimless crowd of clashing elements, which can be seen in a dance of motes of dust whirling in a sunbeam", and highlighted the arrow from order to chaos ("direct cascade" in modern terms) aiming at the final death of the universe (the thermal death in the second law of thermodynamics). Moreover, Lucretius recognized the existence of an alternative "involuntary" arrow from chaos to order ("inverse cascade"). More than two millennia have passed since then until these findings were rigorously expressed and quantified: direct cascade, in the non-stratified 3-dimensional turbulence (Kolmogorov, 1941a,b; 1942); inverse cascades, in the wave turbulence (Zakharov and Filonenko, 1966; Zakharov et al., 1993) and the 2-dimensional hydrodynamic and plasma turbulence (Kraichnan, 1967; Kraichnan and Montgomery, 1980).

The present paper reveals the yet overlooked inverse cascade in convective turbulence, namely, in the buoyancy-generated chaotic plumes, which in fact merge to form larger plumes; whereas mechanical shear-generated chaotic eddies break down to produce smaller ones, thus producing the direct cascade.

Modern vision of stratified turbulence is based on the conventional paradigm generally attributed to Kolmogorov (1941a,b; 1942) with no regard to the fact that his own vision was limited to the non-stratified homogeneous turbulence. So, the paradigm, extended in due time to the stratified turbulence as self-evident, without proof, comprises the following well-known postulates:

- Turbulence develops when the flow instability is strong enough to overtake the resistance of molecular viscosity so that the flow breaks down causing chaotic eddies (Reynolds, 1883)

- These eddies are unstable themselves and also break down to generate smaller unstable eddies, thus causing the direct cascade of the turbulent kinetic energy (TKE) and other properties of turbulence towards smaller scales and eventual viscous dissipation (Richardson, 1920)

- Turbulence results in the downgradient transport of momentum, energy and matter, in other words, it transports these properties in such a way as to level out their spatial heterogeneity; so, the turbulent flux of any quantity is calculated as the product of the

File generated with AMS Word template 1.0

gradient of the transferred quantity and the coefficient of turbulent exchange (Boussinesq, 1897)

- Direct cascade feeds all three shares of TKE making them proportional to each other and yielding isotropic smaller scales, thus assuring the balance between the generation of TKE by velocity shears and its viscous dissipation (Kolmogorov, 1941a,b, 1942)

This vision, particularly as concerns the unstably-stratified turbulence (the subject-matter of the present paper), remained conceptually unchanged until now. For meteorological and oceanographic applications this is not too surprising: all theories based on the above paradigm (Obukhov, 1946; Monin and Obukhov, 1954; Mellor and Yamada, 1982; Canuto et al., 2001) yield plausible quantification of the vertical convective turbulence, while in many applications nothing else is needed. Herewith, experimental evidence against the conventional vision of the unstably stratified turbulence has been gradually accumulated (Zilitinkevich, 1971, 1973; Wyngaard and Coté, 1971; Kader and Yaglom, 1990; Marino et al., 2015; Salesky et al., 2017; Pouquet et al., 2018). In the language of Thomas Kuhn (1962), the state of affairs is called the "crisis of the conventional paradigm".

The present paper demonstrates that the crisis roots in the irrelevance of the conventional paradigm of the unstably stratified turbulence, which arose long ago as a literal paraphrasing of the Kolmogorov vision of the non-stratified homogeneous turbulence. This happened automatically and simply identified buoyant plumes as mechanical eddies.

Instead, the proposed new paradigm, first, declares that plumes principally differ from eddies and merge to form larger ones, thus making the inverse cascade and, second, admits the existence of countergradient and non-gradient turbulent transports prohibited in the conventional paradigm. At the same time, the new paradigm does not contradict the classical Kolmogorov theory formulated for the neutrally-stratified homogeneous turbulence.

## 2. Success and failure of the conventional theory

The conventional and new paradigms of unstably-stratified turbulence are compared below by the example of plane-parallel sheared flow inherent in the atmospheric surface layer. Here, the conventional theory is fair only in the trivial case of so strong shears that mechanical eddies destroy the weak plumes and, thus, violently involve them in the direct cascade. In atmospheric and hydrospheric convective boundary layers such a regime is observed in the near-surface sublayers comprising roughly 10% of the stratified surface layer. Beyond this practically non-stratified sublayer, chaotic plumes produce an inverse cascade culminating at the largest scales in the conversion of convective TKE into kinetic energy of self-organized flow patterns: cells or rolls in the shear-free or sheared convective boundary layers, respectively. Herewith, mechanical turbulence, generated by the mean-flow shears, produces a direct cascade culminating in the viscous dissipation of mechanical TKE into heat at the smallest eddies. So, in the noticeably unstable stratification, horizontal TKE is fully mechanical, while vertical TKE is almost fully convective. This unorthodox picture will be confirmed by experimental data to be featured in the following sections of this work.

The atmospheric surface layer is defined as the near-surface 10% of the boundary layer and provides the most convenient framework for the investigation of the principal nature of stratified turbulence. Indeed, the vertical turbulent flux of the momentum per unit mass, $\boldsymbol{\tau} \equiv (\langle u'w' \rangle, \langle v'w' \rangle)$, and the potential temperature, $F_\theta \equiv \langle w'\theta' \rangle > 0$, are practically independent of the height, $z$, over the surface. Here $u'$, $v'$ and $w'$ are wind velocity fluctuations and $\theta'$ is the fluctuation of potential temperature. Turbulence is fully governed by these fluxes and the buoyancy parameter, $\beta = g/T_0$, where $g$ is the acceleration due to gravity and $T_0$ is the reference value of absolute temperature. From these parameters, one can compose a single length scale, namely, the Obukhov scale:

$$L = |\boldsymbol{\tau}|^{3/2}(\beta F_\theta)^{-1}. \tag{1}$$

So, the dimensionless characteristics of the surface-layer turbulence (normalized using the above governing parameters) become universal functions of $z/L$ (Obukhov, 1946). This makes the surface layer an ideal natural laboratory for the verification and comparison of alternative theories of turbulence.

The energetics of both stably and unstably stratified turbulence is conventionally defined by a single equation quantifying the budget of TKE, $E_K$, principally following the Kolmogorov (1942) vision of the non-stratified turbulence but accounting for the rate of generation of TKE by the buoyancy forces, $\beta F_\theta$, and possible vertical heterogeneity of turbulence. For the steady-state thermally-stratified surface layer, this equation reads (Obukhov, 1946; Monin and Yaglom, 1971):

$$-\tau \frac{\partial U}{\partial z} + \beta F_\theta \equiv \frac{\tau^2}{K_M} + \beta F_\theta = \varepsilon_{K\downarrow} + \frac{\partial F_{EK}}{\partial z}. \tag{2}$$

Here for simplicity, we align the x coordinate along the direction of mean-flow velocity, $U$, so the vertical turbulent flux of the momentum becomes $\tau = \langle u'w' \rangle < 0$; $K_M \equiv -\tau/(\partial U/\partial z)$ is the coefficient of vertical turbulent transport for momentum (often called "eddy viscosity"). The left-hand side of Eq. (2) is the sum of TKE generation rates by the velocity shears and buoyancy forces, respectively; while the right-hand side is the sum of the TKE dissipation rate, $\varepsilon_{K\downarrow}$ (the downward arrow symbolizes the direct cascade), and the vertical divergence of the vertical turbulent flux of TKE, $F_{EK}$, defined as

$$F_{EK} = \langle E'_K w' \rangle + \frac{1}{\rho_0}\langle p'w' \rangle, \tag{3}$$

where $E'_K$ is the fluctuation of TKE, $p'$ is the fluctuation of pressure, $\rho_0$ is a reference value of the air density, and the angle brackets designate statistical averaging.

Kolmogorov (1942) considered the non-stratified turbulence ($F_\theta = 0$) and developed the first theory of the budget of turbulent kinetic energy in this particular case:

- The turbulent time scale is defined as $t_T = E_K/\varepsilon_{K\downarrow}$ (just expressing the unknown $t_T$ via also unknown $\varepsilon_{K\downarrow}$), so that the turbulent length scale is only naturally defined as $l_T = E_K^{1/2} t_T$

- In view of the fact that *in neutral stratification* the turbulent length scale is limited only by distance, *z*, from the solid surface, it should be proportional to this distance, $l_T \sim z$, which immediately defines the time scale: $t_T \sim z/E_K^{1/2}$

- Following Prandtl (1932), the coefficient of vertical turbulent transport for any property, $K_{PROP}$, is defined as proportional to the product of the turbulent length scale, $l_T \sim z$, by the square root of vertical TKE, $E_{KV}^{1/2}$

- Insofar as all shares of TKE have the same origin (the velocity-shear instability): horizontal, $E_{KH} \equiv \langle u'u' \rangle/2 + \langle v'v' \rangle/2$, vertical, $E_{KV} \equiv \langle w'w' \rangle/2$, and total, $E_K \equiv E_{KH} + E_{KV}$, TKE must be similar to each other

This yields the following constructive definitions:

$$\varepsilon_{K\downarrow} = E_K^{3/2}/(C_D z), \tag{4}$$

$$K_{PROP} \sim K_M = C_K E_{KV}^{1/2} z, \tag{5}$$

$$E_{KV} \sim E_{KH} \sim E_K, \tag{6}$$

$$F_{EK} \sim -E_{KV}^{1/2} z \frac{\partial E_K}{\partial z}, \tag{7}$$

where $C_D \approx 4$ and $C_K \approx 0.4$ are dimensionless constants defined empirically (e.g. Monin and Yaglom, 1971; Tampieri, 2017), and $E_{KH}$ is horizontal TKE.

Obukhov (1946) was the first to reveal that the above vision of the TKE budget yields plausible formulations of the vertical TKE and coefficients of vertical turbulent transport not only for neutral or near-neutral stratification:

$$\frac{E_{KV}}{|\tau|} = C \text{ and } \frac{K_M}{|\tau|^{1/2}z} = k, \tag{8}$$

but also for well-pronounced stratification, particularly, for unstable conditions where these formulations read:

$$\frac{E_{KV}}{(\beta F_\theta z)^{2/3}} = C_V \text{ and } \frac{K_M}{z(\beta F_\theta z)^{1/3}} = C_V^{1/2} C_K, \tag{9}$$

where $k \approx 0.4$ is the von Kármán constant; $C \approx 1$, $C_V \approx 1$ and $C_K \approx 0.4$ are well-established empirical dimensionless constants (e.g. Monin and Yaglom, 1971; Garratt, 1992; Stull, 1997; Tampieri, 2017).

It's worth noting that for near-neutral conditions the conventional theory based on equations (6) and (8) implies that the vertical turbulent flux of total TKE is equal to zero: $F_{EK} = 0$, and $\partial F_{EK}/\partial z = 0$ consequently. It is consistent with the local equilibrium between shear production and dissipation rate of TKE in the inertial layer confirmed by measurements and direct numerical simulation data. At the same time, the recent findings by Banerjee and Katul (2013) and Katul et al. (2016) demonstrate the logarithmic scaling for $E_{KH}$ and highlight the deficiency of the conventional theory even for the neutrally stratified surface layer.

Furthermore, along with the plausible Eq. (9), this vision of the TKE budget yields for the unstable stratification ($z > L$) fully erroneous formulations of the following characteristics of turbulence:

vertical flux of TKE

$$F_{EK} \sim -E_{KV}^{\frac{1}{2}} z \frac{\partial E_K}{\partial z} \sim -\beta F_\theta z < 0, \tag{10}$$

horizontal TKE

$$E_{KH} \sim E_{KV} \sim (\beta F_\theta z)^{2/3}, \tag{11}$$

and dissipation rate of TKE

$$\varepsilon_{K\downarrow} = -\tau \frac{\partial U}{\partial z} + \beta F_\theta. \tag{12}$$

Of these results, Eq. (10) follows from the conventional postulate of the only downgradient transport; Eq. (11) follows from the unspoken hypothesis of principal similarity of the buoyancy-generated plumes and shear-generated eddies, which implies that plumes (just like eddies) break down to generate smaller fluid particles spreading in all directions; and Eq. (12) follows from the same hypothesis, which also implies that both shear- and buoyancy-generated types of turbulence make the direct cascade of TKE and therefore are subject to viscous dissipation into heat at the smallest scales. As a result, the conventional theory incorrectly reproduces the contribution of large scale motions to the turbulent transport in the atmospheric surface layer (Salesky and Anderson, 2018, 2020).

## 3. Novel theory

The proposed new paradigm declares the principal difference between the two types of chaotic motions: convective turbulence consisted of vertical buoyant plumes, which merge to produce larger plumes, thus producing inverse cascade; and mechanical turbulence consisted of the shear-generated eddies, which move in all directions and break down to produce smaller eddies, thus making direct cascade. This suggests that total TKE splits into convective TKE, $E_{KC}$, consisting only of the vertical share, and mechanical TKE, $E_{KM}$, consisting of the fully mechanical horizontal share and the mechanical part of the vertical share. Hence, the TKE budget equation, Eq. (2), splits into separate equations:

$$\beta F_\theta = \frac{\partial F_{EKC}}{\partial z} \equiv \varepsilon_{K\uparrow}, \qquad (13)$$

$$-\tau \frac{\partial U}{\partial z} = \frac{\tau^2}{K_M} = \varepsilon_{K\downarrow}, \qquad (14)$$

where $F_{EKC}$ is the vertical flux of the convective TKE. Since the contribution of the mechanical turbulence to the vertical turbulent flux of the vertical TKE should be negligible, we approximate $F_{EKC}$ with $F_{EKV} \equiv \frac{1}{2}\langle w'^3\rangle + \frac{1}{\rho_0}\langle p'w'\rangle$. This approximation is validated below (see Figure 2), proving that the vertical flux of the mechanical TKE is almost negligible and $F_{EKC}$ practically coincides with the vertical flux of the total TKE, $F_{EK}$.

According to Eq. (13), the rate of *production* of convective TKE, $\beta F_\theta$, is balanced by the term $\partial F_{EKC}/\partial z$, which, therefore, must signify a rate of its *consumption*. The only candidate for this role is the rate of the yet overlooked *conversion of convective TKE into kinetic energy of large-scale self-organized convective flow patterns* (Elperin et al., 2002, 2006; Zilitinkevich et al., 2006). If so, the term $F_{EKC} \approx F_{EK}$, traditionally interpreted as the downgradient vertical flux transporting TKE downward, is, in fact, the *countergradient* (oriented upwards) flux in the physical space and, at the same time, the inverse flux (from smaller to larger scales) over the spectrum towards the conversion of TKE into kinetic energy of self-organized flow patterns. Experimental data shown in Figure 2 confirm this fully non-orthodox vision.

Notably, the balance between the rate of the TKE production by buoyancy forces and the vertical gradient of the vertical turbulent flux of TKE was long ago revealed empirically by Wyngaard and Coté (1971). However, the general recognition of the conventional paradigm as applicable to any stratified turbulence was unshakable, and this empirical discovery went almost unnoticed. In the light of the new paradigm, it means nothing but direct empirical evidence of the separate budget of convective TKE, Eq. (13) that, in turn, testifies the separate budget of mechanical TKE, Eq. (14).

The rate of viscous dissipation of mechanical TKE, $\varepsilon_{K\downarrow}$, and the rate of conversion of convective TKE, $\varepsilon_{K\uparrow}$, both quantify the TKE consumption, with the only difference that dissipation consumes the smallest eddies while conversion, the largest plumes. Then, reshaping the Kolmogorov interpretation of the dissipation rate of mechanical turbulence, Eq. (4), for the rates of both, dissipation of mechanical TKE and conversion of convective TKE, these two processes are on equal terms quantified as the ratios of the appropriate energies, $E_{KM}$ or $E_{KC}$, by the common turbulent time scale, $t_T \sim z/E_K^{1/2} \approx z/E_{KC}^{1/2}$. This yields the following novel formulations:

$$\varepsilon_{K\uparrow} = \frac{E_{KC}^{3/2}}{C_\uparrow z}, \tag{15}$$

$$\varepsilon_{K\downarrow} = \frac{E_{KM} E_{KC}^{1/2}}{C_\downarrow z}, \tag{16}$$

where $C_\uparrow$ and $C_\downarrow$ are dimensionless constants to be determined empirically (the upward and downward arrows symbolize the inverse and direct cascades, respectively).

Notably, the idea of turbulent length scale as proportional to the height over the surface, $l_T \sim z$, underlying the Kolmogorov concept of dissipation, Eq. (4), is equally relevant to conversion and unstable stratification as the latter does not impose any additional limitations on the vertical turbulent length scale.

We shall emphasise that Eqs. (15-16) may not hold for near-neutral conditions: the exact area of validity and the transition to the neutral limit in terms of $z/L$ are to be determined separately. It would require high-resolution numerical simulation data tailored specifically to resolve this issue.

## 4. Breaking the dead-end of conventional theories

Now it becomes clear how it happens that the principally erroneous conventional theory yielded plausible formulations of the vertical TKE and the coefficient of vertical turbulent transport of momentum specified by Eq. (9). Indeed, insofar as the share of mechanical turbulence in total TKE is small, the *physically erroneous conventional vision of the total TKE budget*, Eqs. (4-7), as its total production balanced by viscous dissipation serves as the lucky approximation of the *real budget of convective TKE,* Eqs. (13, 15), namely, as its production by the buoyancy forces balanced by its conversion into kinetic energy of self-organized flow patterns. So, in spite of the principal difference between the conventional and new visions/theories, both yield mathematically the same expressions of the parameters in question. Herewith, conventional theory yields fully erroneous Eqs. (10-12) specifying all the remaining characteristics of turbulence.

Contrastingly, the novel theory yields, besides Eq. (9), the true formulations of these characteristics:

vertical flux of the vertical TKE, $F_{EKV}$, oriented upward (counter gradient):

$$\frac{F_{EKC}}{|\tau|^{3/2}} \approx \frac{F_{EKV}}{|\tau|^{3/2}} \approx \frac{F_{EK}}{|\tau|^{3/2}} = \frac{C_V^{3/2}}{C_\uparrow} \frac{z}{L} > 0, \qquad (17)$$

fully mechanical horizontal TKE, $E_{KH}$, decreasing with $z/L$:

$$\frac{E_{KH}}{|\tau|} = C_H \left(\frac{z}{L}\right)^{-2/3}, \qquad (18)$$

the rate of the viscous dissipation of TKE, $\varepsilon_{K\downarrow}$, balancing only the rate of its generation by the mean velocity shear and having nothing to do with the buoyancy-generated turbulence:

$$\frac{\varepsilon_{K\downarrow}z}{|\tau|^{3/2}} = \frac{1}{C_V^{1/3}C_K}\left(\frac{z}{L}\right)^{-1/3}. \tag{19}$$

Here the dimensionless constants $C_V \approx 1$ and $C_K \approx 0.4$ are well known from the empirical validation of Eq. (9), common for the conventional and novel theories, while the new constants $C_\uparrow = C_V^{3/2} \approx 1$ and $C_H \approx 8.4$ are defined from the data shown in Figures 2 and 3.

Eq. (18) predicts that fully mechanical horizontal TKE, $E_{KH}$, increases approaching the surface. This dependence on height has been observed both in free shear convective boundary layers and neutrally-stratified flows. The logarithmic dependence for the latter was shown by, e.g., Banerjee and Katul (2013) and Katul et al. (2016) and confirmed by DNS data. The LES data (Abkar and Moin, 2016) of the convective boundary layer also supports this finding and matches observational data. We expect this relation to hold for all intermediate regimes of sheared-convective turbulence.

It should be noted that the coefficient of vertical turbulent exchange for momentum $K_M$ is derived rigorously from the budget equation for the vertical turbulent flux of momentum. But in the proposed framework this equation splits into the pair of equations for its convective and mechanical parts similarly to splitting of the convective and mechanical TKE budgets into Eqs. (13-14), so $K_M$ should be rederived as well. The authors leave this derivation for a separate paper.

## 5. Empirical validation

Experimental data needed for validation of Eqs. (17-19) are obtained from meteorological observations at the Eureka station located in the Canadian territory of Nunavut in the conditions of long-lived convective boundary layers typical of the Arctic summer. Here, the permanent warming of the layer from the surface is balanced by the permanent pumping of colder air into the layer via the general-circulation mechanisms. Such balance yields the quasi-stationary regime of turbulence assuring more accurate detection of turbulent energies and fluxes than is possible in the short-lived evolving convective layers typical of mid-latitudes. Herewith, no principal contradictions were found between the available data from observations at mid or low latitudes and the more certain data from Eureka (Grachev et al., 2018).

Turbulent energies and fluxes were calculated directly from the measured velocity and temperature fluctuations. The pressure-velocity correlation term could not be evaluated directly using the available measurement data, the only estimate could be retrieved from the TKE budget equation. Since all other terms are known and the overall residual of the balance equation never exceeds 10%, the pressure-velocity correlation term is neglected. We attribute the relative unimportance of this term to a narrow time averaging applied (the known TKE budget terms were evaluated as a result of time averaging over 100 s intervals). In the light of the new paradigm, this result is quite understandable: in the unstably stratified turbulence, breaking eddies causing the pressure fluctuations to make only little contribution to TKE for the time scales considered, while buoyant plumes dominating this turbulence do not cause pronounced pressure fluctuations.

Notably, the TKE dissipation rate is not measured directly. For a long time, the only possibility was to retrieve it from the TKE budget equation as the residual term, provided that all other terms are measured. However, for neutral and stratified turbulence this equation includes the unmeasurable pressure-velocity correlation term. This makes the method rather uncertain. Comparatively recently, a constructive method of retrieving the dissipation rate from the measured TKE spectra was proposed (Pearson et al., 2002). It employs the Kolmogorov idea that the shape of the TKE spectrum in the inertial interval is fully controlled by the dissipation rate; so the latter can be calculated from the measured spectra. In neutral and stable stratifications the method does not raise questions: the TKE spectrum exhibit just one inertial interval in its high-frequency part. However, in unstable stratification the TKE spectra exhibit two inertial intervals – one at higher frequencies and another at lower frequencies (e.g. Kader and Yaglom, 1991; Glazunov and Dymnikov, 2013; Banerjee et al., 2015). Until the present, the second interval remained mysterious. The new paradigm only naturally explains it as the manifestation of the inverse cascade in convective turbulence towards its conversion into the large-scale self-organised structures. So, the shape of the spectrum in this low-frequency interval is controlled by the rate of conversion of convective TKE into kinetic energy of self-organised structures, just as the spectrum in the high-frequency interval is controlled by the dissipation rate (see Figure 1). According to Eq. (13) this rate, $\varepsilon_{K\uparrow}$, is balanced by the rate of production of convective TKE quantified by the vertical turbulent flux of buoyancy, $\beta F_\theta$. So, the above method gives the TKE conversion rate, $\varepsilon_{K\uparrow} = \beta F_\theta$, when applied to the low-frequency inertial interval, and the TKE dissipation rate, $\varepsilon_{K\downarrow} = -\tau \frac{\partial U}{\partial z}$, when applied to the high-frequency interval. The empirical data on dissipation shown in Figure 4 were retrieved by this method from the high-frequency inertial interval utilizing the value of 0.55 for the Kolmogorov constant used by Grachev et al. (2015) for the energy spectrum of the longitudinal velocity component.

Until now, the different nature of the two inertial intervals in the TKE spectra remained unknown. As a result, the low-frequency interval was often used for retrieving the dissipation rate all the more confidently that such procedure gave the results consistent with the conventional theory. It is this mistake that Kader and Yaglom (1990) made in their comprehensive experimental investigation of various statistical moments of the surface-layer turbulence. So, the imaginary rate of the TKE dissipation retrieved in this very informative study is in fact the rate of the TKE conversion into kinetic energy of self-organized structures, which exactly balances the rate of the TKE generation – in strict accordance with the new theory. This confusion, only natural for that time, deserves notice to warn modern users of the method against such mistakes.

Figure 2 clearly demonstrates the countergradient vertical transport of TKE predicted by the new theory and yields a certain empirical estimate of the dimensionless constant $C_\uparrow \approx C_V^{3/2} \approx 1$. This reveals the perfect balance between the rate of generation of convective TKE, $\beta F_\theta$, and the rate of its consumption, $\varepsilon_{K\uparrow} = \partial F_{EK}/\partial z \approx \partial F_{EKV}/\partial z$, thus proving the separate budget of convective TKE, Eq. (13), and the entailing separate budget of mechanical TKE, Eq. (14). The right panel of figure 2 demonstrates that the vertical fluxes of total and vertical TKE, $F_{EK}$ and $F_{EKV}$, practically coincide at the large enough values of $z/L$, which means that mechanical turbulence practically does not contribute to the vertical flux of TKE.

Figure 3 confirms the predicted opposing behaviour of horizontal TKE, $E_{KH}/|\tau| = C_H(z/L)^{-2/3}$, decreasing with $z/L$ and vertical TKE, $E_{KV}/|\tau| = C_V(z/L)^{2/3}$, increasing with $z/L$. It yields a certain empirical estimate of the constant $C_H = 8.4$, and clearly demonstrates the irrelevance of the conventional paradigm postulating the only direct cascade and the entailing similarity of all shares of TKE.

Figure 4 provides direct empirical evidence of the novel vision of the TKE dissipation rate, $\varepsilon_{K\downarrow}$, defined after Eq. (14) or, equivalently, after Eq. (19). Experimental data on dissipation are retrieved (as explained above) from the high-frequency inertial intervals in the TKE spectra. The figure shows the perfect balance between the TKE dissipation and the easily measurable TKE generation by the velocity shears, Eq. (14), exactly complementing the separate budget of convective TKE, Eq. (13).

File generated with AMS Word template 1.0

The theoretical solid blue curve showing $\varepsilon_{K\downarrow}$ in the noticeably unstable stratification ($z > L$) agrees with experimental data without any fitting. It is all the more illustrative that the curve is plotted using conventional values of the constants $C_V \approx 1$ and $C_K \approx 0.4$ obtained from the independent empirical validation of Eq. (9). To complete the picture, the dotted blue line shows the extension of novel theory beyond the area of its relevance – to the practically non-stratified sub-layer $z < L$. The conventional theory, based on the unproven claim that the shear- and buoyancy-generated TKE are both subjected to viscous dissipation at small scales, is shown by the solid black line inside the narrow mechanical sublayer (where it holds true) and by the dotted black line outside this layer (where it fails). As seen from this figure, the conventional theory overestimates the dissipation, $\varepsilon_{K\downarrow}$, up to an order of magnitude. So, the figure clearly demonstrates that in the pronounced unstable stratification the lion share of TKE is not subjected to viscous dissipation.

## 6. Conclusions

This paper revises the conventional paradigm of turbulence as irrelevant to the unstably-stratified fluid flows characterised by the two fully different mechanisms of generation of turbulence: the convective instability and the velocity-shear instability. The proposed new paradigm is declared, demonstrated by the example of turbulence energetics in the unstably stratified atmospheric surface layer, and proved experimentally. Its key point is the principally different nature of the two types of turbulence conventionally considered as similar: buoyancy-generated plumes and shear-generated eddies. In contrast to eddies, breaking down into smaller ones and thus chaotically spreading in three dimensions, plumes adhere to the buoyancy-oriented vertical motions and do not break down but merge into larger plumes. This implies coexisting of the two principally different types of chaotic motions:

- The familiar mechanical turbulence, that produces the direct cascade from the larger to smaller scales culminating at minimal scales in the viscous dissipation of TKE into heat

- Completely different convective turbulence that makes the inverse cascade from the smaller to larger scales culminating at maximal scales in its conversion into large-scale self-organized flow patterns

This non-orthodox vision of turbulence admits the two principal arrows:

- "Chaos out of order" (paraphrase of the thermodynamic concept of thermal death) discovered by Richardson (1920) and utilized in the Kolmogorov paradigm of the non-stratified turbulence

- "Order out of chaos", conceptually analogous to the Nietzsche "creative chaos", discovered by Prigogine as inherent to the self-organization, particularly in life systems (Prigogine and Stengers, 1984), and utilized in the proposed new paradigm of the unstably-stratified turbulence

The new paradigm, when it has already formed, seems self-evident and almost trivial. It may seem unclear why the inverse cascade in convective turbulence has gone unnoticed for so long. This is all the more amazing that such a cascade could not help but come to mind being the only reasonable explanation of the energy supply for the self-organized convective flow patterns inherent in numerous natural phenomena from convective boundary layers to stellar convection. However, the hypnosis of the conventional paradigm happened to be so strong that its revision took half a century from the first virtually unheard signals of its fallacy (Zilitinkevich, 1971, 1973; Wyngaard and Coté, 1971).

*Acknowledgments.*

The authors acknowledge support from the Academy of Finland Flagship funding (grant no. 337549); collection and processing of the measurement data used for empirical validation were supported by the Ministry of Science and Higher Education of Russia (contract No. 14.W0331.0006 and agreement No. 075-15-2019-1621); data analysis and interpretation was partially supported by Russian Science Foundation, grant No. 21-71-30023.

*Data Availability Statement.*

The data can be accessed at http://doi.org/10.23728/b2share.bb8964ff899c4711a0e8875b87ab2800 (Kadantsev and Repina, 2020).

FIGURES

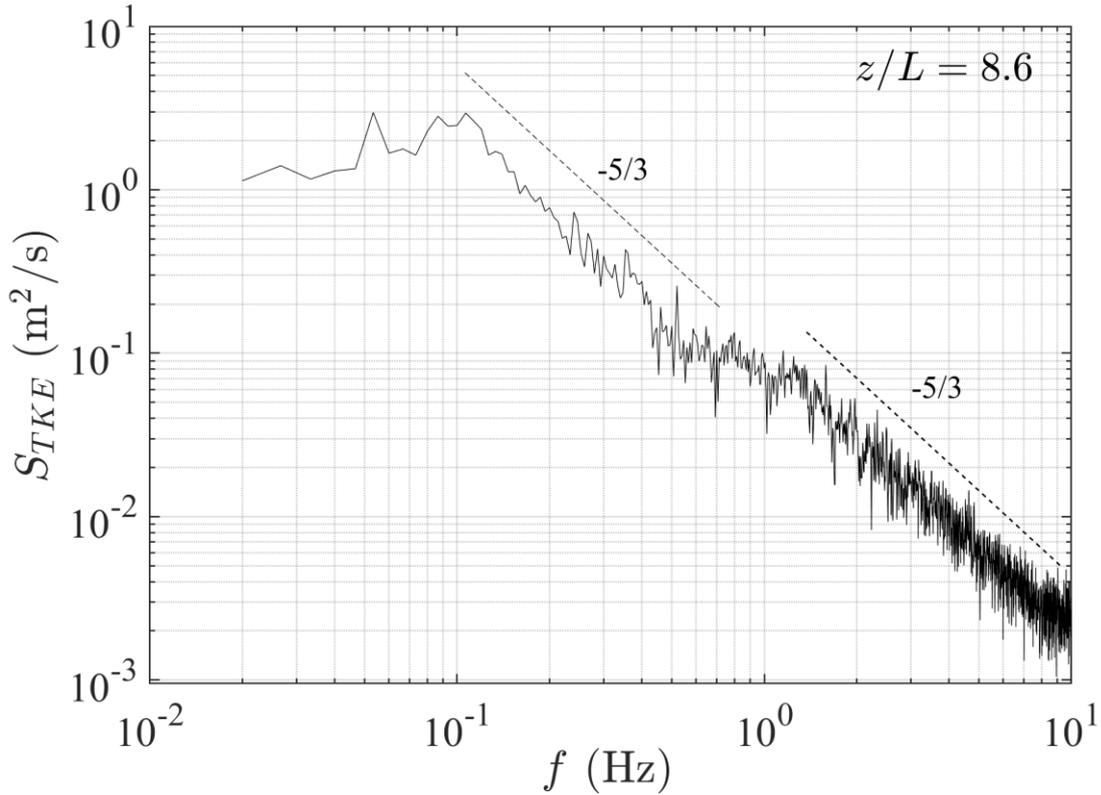

Figure 1: Typical spectrum, $S_{TKE}$, of turbulent kinetic energy (TKE) measured at the Eureka research station (8 m above the surface; Obukhov length, $L \sim 1$ m; 22.06.2012, 18:00-18:30 GMT). The two inertial intervals in the spectrum characterised by the -5/3 power law are clearly seen. The low-frequency interval, $0.15 < f(\text{Hz}) < 0.7$, is controlled by the rate of conversion of convective TKE into kinetic energy of self-organised structures, $\varepsilon_{K\uparrow}$, balanced by the TKE buoyant production, $\beta F_\theta$; whereas the high-frequency interval, $1.5 < f(\text{Hz}) < 10$, is controlled by the TKE dissipation rate, $\varepsilon_{K\downarrow}$, balanced by the rate of its generation by shear, $-\tau \frac{\partial U}{\partial z}$, and just used to retrieve $\varepsilon_{K\downarrow}$.

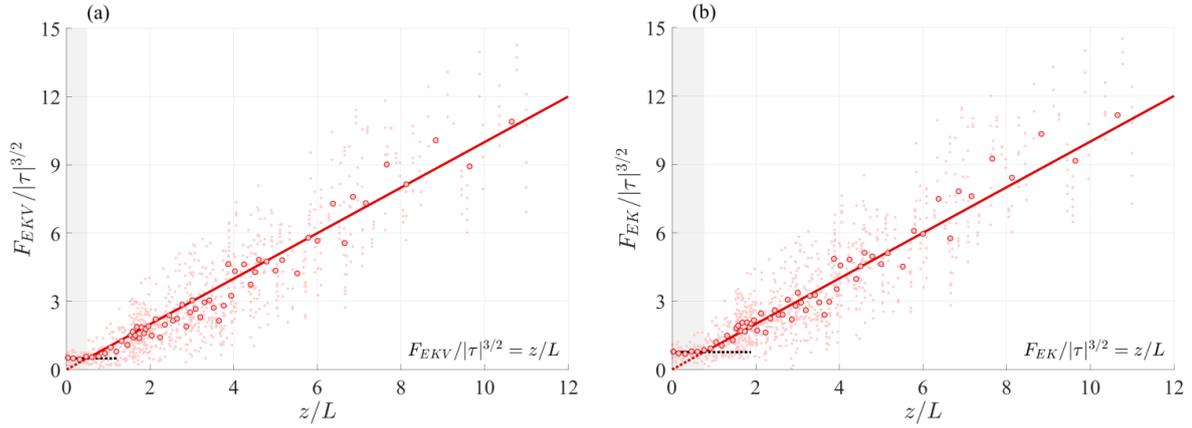

Figure 2: The countergradient nature of vertical turbulent fluxes of the vertical and total TKE, $F_{EKV}$ and $F_{EK}$. Panels (a) and (b) show dimensionless fluxes, $F_{EKV}/|\tau|^{3/2}$ and $F_{EK}/|\tau|^{3/2}$, versus dimensionless height, $z/L$, quantifying the effect of stratification on turbulence. In full agreement with novel theory, in the noticeably unstable stratification ($z > L$), solid lines plotted after Eq. (17) with empirical constant $C_\uparrow = C_V^{3/2} = 1$ coincide with the medians of empirical data. So, panel (a) yields $F_{EKV} = \beta F_\theta z$, which directly confirms the separate budget of convective TKE, Eq. (13). The very close similarity between the two panels reveals negligence of the contribution of mechanical turbulence to the fluxes under consideration. In the almost neutrally stratified sublayer, $z < L$, the ratio $F_{EKV}/|\tau|^{3/2}$ is constant with height in accordance with classical Kolmogorov theory, whereas for $z > L$ this theory is inapplicable. In this and further figures, the dotted parts of red lines show the extension of the novel theory to the sublayer $z < L$, whereas the dotted parts of black lines show the extension of the conventional theory beyond this sublayer. Small dots show the data immediately retrieved from observations, heavy dots show the ensemble means.

File generated with AMS Word template 1.0

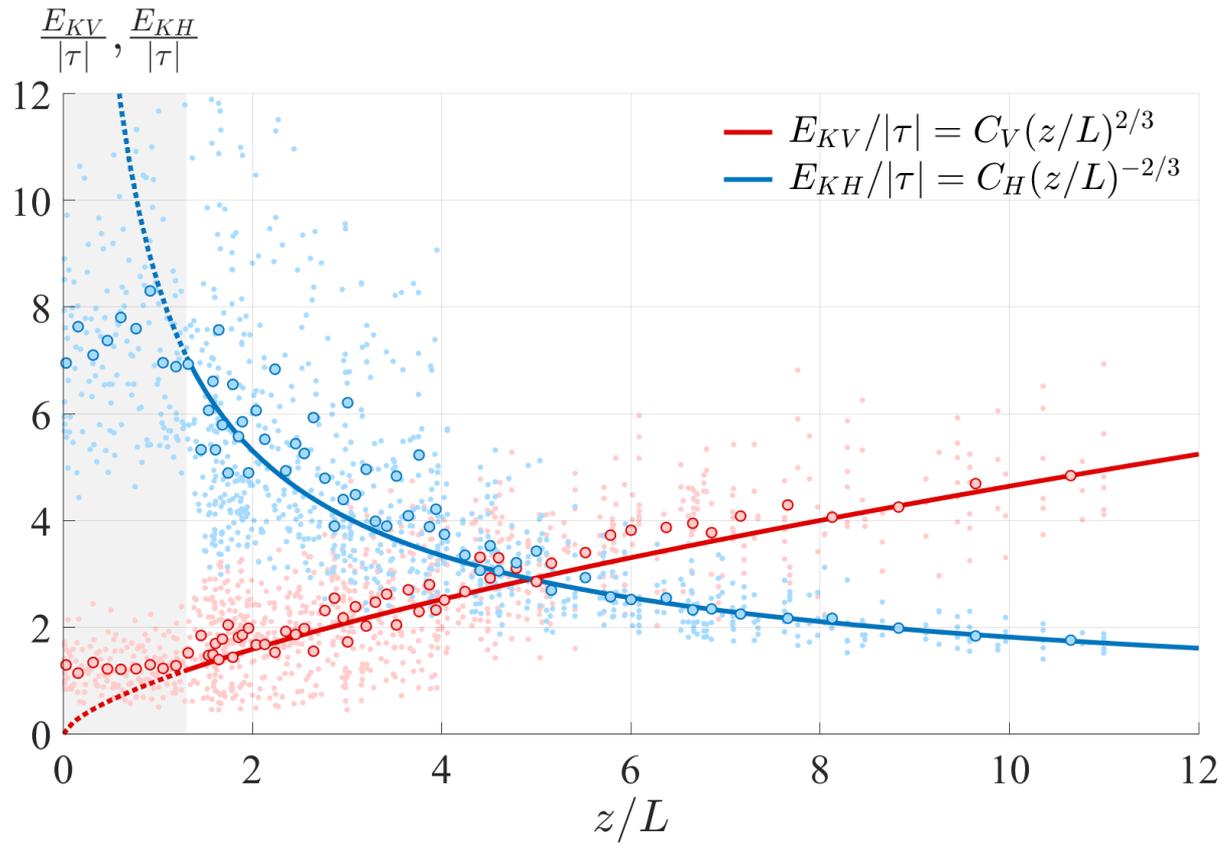

Figure 3: Comparison of dimensionless vertical profiles of the fully mechanical horizontal TKE, $E_{KH}$, and the dominantly convective vertical TKE, $E_{KV}$, shown by the blue and red points/curves, respectively. The curves show $E_{KH}$ after non-orthodox Eq. (18) with $C_H = 8.4$, and $E_{KV}$ after Eq. (9) with $C_V = 1$ resulting from both novel and conventional theory. As seen in the figure, the conventional definition of horizontal TKE by Eq. (11) has nothing in common with reality.

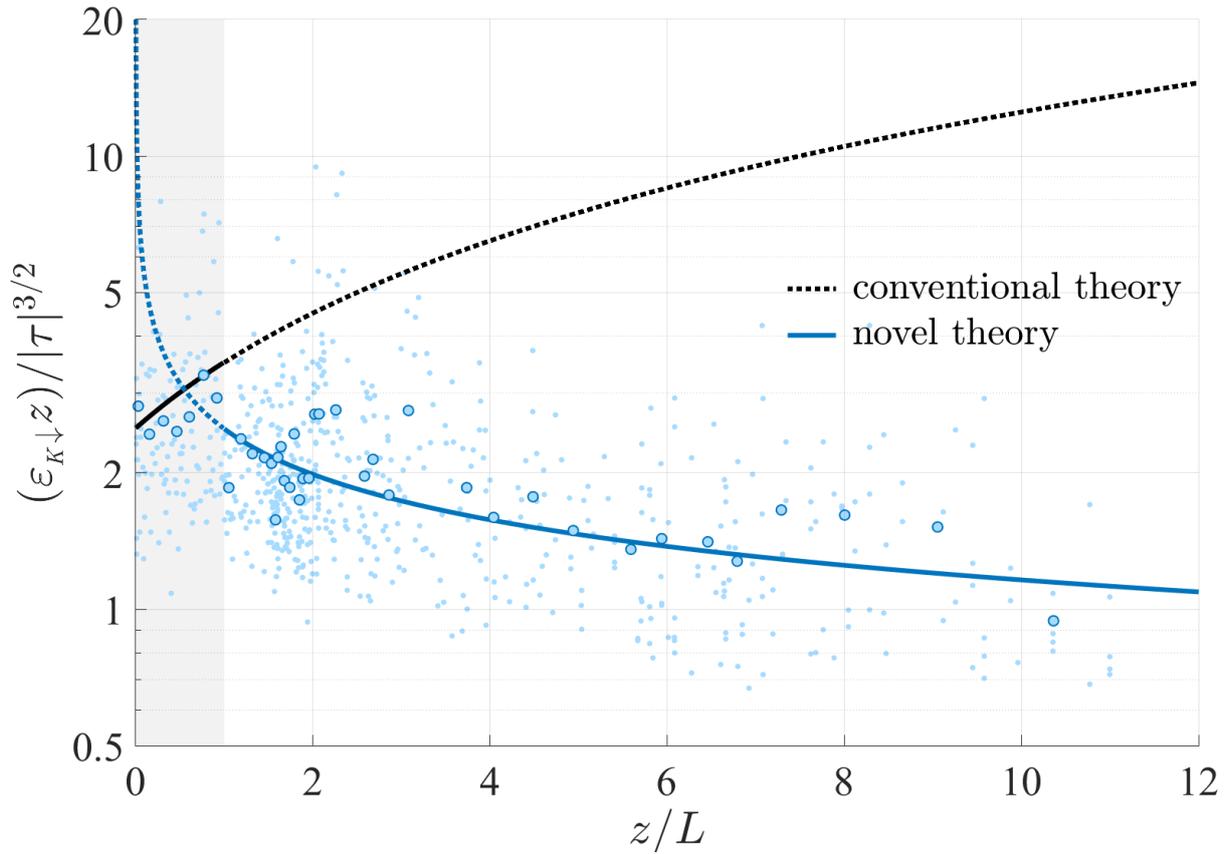

Figure 4: Novel and conventional visions of the TKE dissipation rate, $\varepsilon_{K\downarrow}$, in unstably stratified turbulence. The non-orthodox Eq. (19), defining the dissipation, $\varepsilon_{K\downarrow}$, as inherent only in mechanical turbulence, is shown at $z > L$ by the solid blue line, and at $z < L$ (in the almost neutral stratification) by the dotted blue line. The conventional theory, assuming that both shear- and buoyancy-generated TKE are subjected to viscous dissipation, is shown at $z < L$ by the solid black line and at $z > L$ (beyond the area of its validity) by the dotted black line. Blue dots show empirical values of $\varepsilon_{K\downarrow}$ retrieved from the low-frequency (large-scale) inertial intervals of the measured spectra of TKE.